\documentclass[10pt,aps,pre,twocolumn,superscriptaddress,floatfix,showpacs,longbibliography]{revtex4-1}
\usepackage[utf8]{inputenc}
\usepackage{graphicx}
\usepackage{tabularx}
\usepackage[usenames,dvipsnames,table]{xcolor}
\usepackage[version=3]{mhchem}
\usepackage{braket}
\usepackage{upgreek}
\usepackage{hyperref}
\usepackage{amsmath, amssymb}
\usepackage[normalem]{ulem}
\usepackage{soul}

\definecolor{mark}{rgb}{0.85, 0.9, 1}
\definecolor{rred}{HTML}{CB4154}

\hypersetup{hidelinks}
\sethlcolor{mark}
\newcolumntype{Y}{>{\centering\arraybackslash}X}

\begin{document}

\title{Estimating classical mutual information between quantum subsystems with neural networks}
\author{D. A.  Konyshev}
\affiliation{ Theoretical Physics and Applied Mathematics Department, Ural Federal University, Mira Str. 19, 620002 Ekaterinburg, Russia}
\author{V. V. Mazurenko}
\affiliation{ Theoretical Physics and Applied Mathematics Department, Ural Federal University, Mira Str. 19, 620002 Ekaterinburg, Russia}

\date{\today} 

\begin{abstract}
Characterizing correlations in a quantum system on the basis of the results of the projective measurements can be performed with different means including the calculation of the classical mutual information. Generally, estimating such information-entropy-based quantities requires having complete statistics of the system's states. Here we explore the possibility to reconstruct the classical mutual information and specific entropy of a quantum system with neural network approach on the basis of limited number of projective measurements. As a prominent example we consider the antiferromagnetic quantum Ising model in transverse and longitudinal magnetic fields which is in demand in both condensed matter physics and quantum computing. We show that the neural network approach gives reliable estimates of the classical mutual information even in the case of paramagnetic wave functions delocalized in the state space. In addition, the phase diagram of the considered quantum system is reconstructed with a special focus on discriminating various types of disordered states. 
\end{abstract}
\maketitle

\section*{Introduction}
The first step when describing a complex multi-component system in physics, biology, machine learning and other fields is to reveal and quantify correlations between its different parts. Understanding correlations structure facilitates comparing a given system to others and constructing models to simulate its properties. However, there is no a unique way to describe correlations. Generally, it can be done with different means and at different levels. For instance, one of the widely used approaches is based on the calculation of two-point correlation functions which belong to the family of Ursell connected correlation functions \cite{Ursell}. This enables theoretical description of collective behavior in systems of completely different origin and scales, for instance bird flocks \cite{Parisi}, spin glasses \cite{glasses}, optical interferometers \cite{boson11, boson12}, equilibrium \cite{Imada, qskyrmion} and non-equilibrium \cite{DTC, DTCour} quantum systems, and many others. 

Another approach for exploring correlations in a system is provided with the information theory developed by Shannon \cite{Shannon}. Here one is to calculate a mutual information (MI) that quantifies the amount of information about a subsystem that becomes available when one measures another one.
Using MI is a standard tool for characterizing numerous physical \cite{Bialek, Vicsek, PT_MINE_Kondo, MI_Area_law} and artificial systems \cite{ML_Review, NN_understanding, GANs}. Searching for connection between the mutual information and two-point correlation functions explored in Refs.\onlinecite{Matsuda, QI_Sf_Area_laws} reveals a number of important results. For instance, comparison of the area laws for quantum and classical systems shows that MI for quantum states could be larger than that for classical ones. On the other hand, while both MI and two-point correlators are sensitive to phase transitions \cite{Matsuda, Vicsek, PT_MI}, they could demonstrate different behavior near the critical points \cite{MI_spin_models}. 
In addition, finding relationship between classical MI and entanglement entropy discussed in Refs.\onlinecite{Sf_MI,Sf_MI_,Entanglement_Many_Body_systems} facilitates analysis of the experimental data.
Thus, exploring physical models using information theory methods looks rather attractive.

The direct calculation of mutual information is complicated by the fact that in most cases one deals not with known distributions but with a limited number of examples subordinate to their statistics. A large number of parameterized methods have been developed for providing numerical estimates of mutual information \cite{MINE, MI_estimators, MI_errors}. Their goal is to maximize the lower bound of mutual information with the hope that this bound will be close to the exact MI value \cite{MI_estimators}. However, none of the methods does not suffer from different limitations and normally depends on a certain number of samples \cite{MI_errors}. In addition, it has been proved that any distribution-free high-confidence lower bound will not exceed the logarithm of the number of examples $n$ on which this estimated bound is based. This becomes crucial when approximating high-dimensional random variables with MI taking large values (hundreds of bits) \cite{MI_errors}.
 
In this paper we demonstrate utility of the neural networks for estimating classical mutual information between quantum subsystems with a limited number of bitstrings. More specifically we use a mutual information neural estimator (MINE) \cite{MINE} for exploring correlations in the antiferromagnetic quantum Ising model in transverse and longitudinal fields that represents a special interest in condensed matter physics and quantum computing \cite{Exact_Sf,MI_spin_models_}. First, the accuracy of the neural network estimates of MI is examined by the example of the antiferromagnetic and paramagnetic phases. The advantage of this approach is demonstrated through comparing with the results of brute-force method of calculating MI in which the probabilities are reconstructed directly from available bitstrings. Then, we discuss the correspondence between the values of the classical mutual information and entanglement entropy in a wide range of magnetic fields. This facilitates quantifying quantum entanglement directly from a limited number of measurements in single basis avoiding an expensive and approximate reconstruction of the density matrices. Finally, on the basis of the specific entropy calculated with the neural network approach we reconstruct the phase diagram of the Ising model and analyze the transitional areas between different phases.  

\section*{Methods}
The optimization algorithm MINE \cite{MINE} we use to estimate MI is based on Kullback-Leibler divergence between the probability distributions of the states of two subsystems, $A$ and $B$, under consideration:
 \begin{equation}\label{MI}
    M\left(A,B\right){\equiv}D_{\mathit{KL}}\left(P_{\mathit{AB}}{\parallel}P_{A\times B}\right),    
 \end{equation}
where $P_ {\mathit{AB}}$ is the total distribution of the states of the system in question, $P_{A\times B}$ is the product of the marginal distributions $P_A=\int _\mathcal{B}^{} dP_{\mathit{AB}}$ and $P_B=\int _\mathcal{A}^{}dP_{\mathit{AB}}$. Applying the Donsker-Varadan theorem, known as the compression lemma \cite{theorem},
   \begin{equation}\label{D_KL}
    D_{\mathit{KL}}\left(P{\parallel}Q\right){\geq}{\sup _{{T:P\times Q\rightarrow \mathbb{R}}}\left[\mathbb{E}\left[T\right]_P-\log
    \left(\mathbb{E}\left[e^T\right]_Q\right)\right]},
   \end{equation}
to Eq.\ref{MI} one gets the following expression for the MI estimate: 
   \begin{equation}\label{MI_D_KL}
     M_{\boldsymbol{\theta}}(A,B)=\sup_{\boldsymbol{\theta} \in \boldsymbol{\Theta}}\left[\left\langle f_{\boldsymbol \theta }\left(A,B\right)\right\rangle _{P_{\mathit{AB}}}-\log
     \left \langle e^{f_{\boldsymbol \theta }\left(A,B\right)}\right\rangle _{P_{A\times B}}\right].
    \end{equation}
Here one is to find a suitable parameterized function $f_{\boldsymbol \theta}$, optimized in such a way that the expression Eq.\ref{MI_D_KL} reaches its maximum. As such a function, one can choose a neural network that has universal approximating properties \cite{Hornik}. In this case the vector $\boldsymbol \theta$ contains the weights of the neural network.

\begin{figure*}[]
    \centering
    \includegraphics[width=1\textwidth]{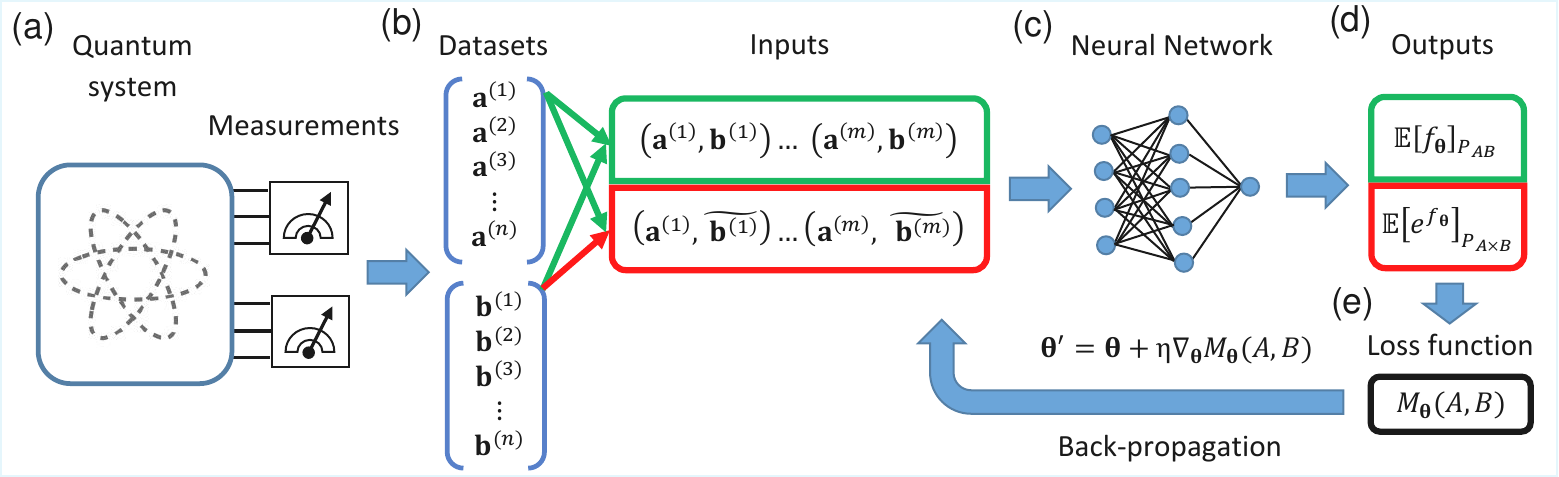}
    \caption{Schematic representation of the procedure for estimating classical mutual information between quantum subsystems with the MINE approach. \textbf{(a)} Generation of data (bitstrings) describing a certain physical system. \textbf{(b)} Two training sets (batches) are created, each of which corresponds to its own probability distribution (joint or product of the marginals) of states. \textbf{(c} and \textbf{d)} Scheme of a neural network used to calculate the average values required for Eq.\ref{MI_D_KL}. \textbf{(e)} By using back-propagation method, the derivatives of the loss function are calculated and, then, network weights are renewed. $\eta$ stands for the learning rate. The iterative procedure is repeated until the convergence is reached.}\label{FIG1} 
\end{figure*}

The MINE procedure for finding classical mutual information between quantum subsystems denoted as $A$ and $B$ includes the following steps that are schematically presented in Fig.\ref{FIG1}. First, $n$ states (bitstrings), ${\bf a}^{i}$ and ${\bf b}^{i}$ ($i=1...n$) of a given quantum system are sequentially generated with projective measurements of both subsystems $A$ and $B$ in parallel (Fig.\ref{FIG1}\textbf{a}) in the computational basis. Then, to prepare a training set one combines pairs of bitstrings in the form $({\bf a}, {\bf b})$ corresponding to the joint probability distribution $P_{AB}$, and $({\bf a}, \widetilde{\bf b})$ corresponding to the products of the marginal distributions $P_{A \times B} = P_{A} \times P_{B}$ (they are shown in Fig.\ref{FIG1} with green and red frames, respectively). The tilde sign $\langle \sim \rangle$ denoted  the configurations of the subsystem $B$  (the red arrow)  indicates their random ordering when combining with the states of the subsystem $A$. Accordingly, subsystem`s configurations without this sign indicate samples in the original order (green arrows) as it was obtained from the projective measurements of the whole system. The composed pairs of the configurations are divided into small batches (Fig.\ref{FIG1} {\bf b}), and used as input of a neural network (Fig.\ref{FIG1}\textbf{c}) that corresponds to the parameterized function $f_{\boldsymbol \theta}$ in Eq.\ref{MI_D_KL}. The average values of $\mathbb{E}[f_{\boldsymbol \theta}]_{P_{AB}}$ and $\mathbb{E}[e^{f_{\boldsymbol \theta}}]_{P_{A\times B}}$ are estimated with neural network and used to approximate the classical mutual information, $M_{\boldsymbol{\theta}}(A,B)$ in according with Eq.(\ref{MI_D_KL}), which is the loss function in that case. The neural network parameters are updated using the back-propagation algorithm (Fig.\ref{FIG1}\textbf{e}) which computes the gradient $\nabla_{\boldsymbol{\theta}} M_{\boldsymbol{\theta}}(A,B)$ to maximize the mutual information. The described procedure is repeated until the loss function stops improving and $M_{\boldsymbol{\theta}} (A,B)$ saturates. 

As for the type of neural network architecture that can be used in MINE approach, it can be different and depends on the properties of the system in question. As demonstrated in previous works, fully connected FFN \cite{PT_MINE_Kondo}, convolutional CNN \cite{MICE} and other modified neural networks \cite{RSMINE, RSMINE_, NIS} provide reliable estimates of $M(A,B)$ for different physical systems. Since our study assumes utilizing one-dimensional input data, we use a fully-connected neural network with the ReLU activation function and the SGD optimizer. More information on the details of the neural network learning and post-processing can be found in Appendix \textbf{A}. 

The MINE method can also be used as an integral part for other numerical approaches that estimate information-entropy-based quantities \cite{MICE, RSMINE, RSMINE_}. For example, the MICE algorithm \cite{MICE} facilitates determination of the specific entropy that can be used to construct phase diagrams of different physical systems and determine the nature of critical boundaries. The idea behind this method is to define the information entropy in terms of the mutual information of its constituent subsystems, which is expressed in the form of an expression for the joint entropy \cite{Shannon}
 \begin{equation}\label{MI_def}
    H(A,B) = H(A) + H(B)- M(A,B).   
    \end{equation} 
Here $H (A)=-\sum_i p_i \log_2 p_i$ is the Shannon information entropy describing a subsystem $A$. 
An additional condition assuming the translational symmetry of the target system determines the statistical indistinguishability of subsystems of equal sizes, $H(A)=H(B)$. All possible locations of a subsystem are equivalent for collecting data about it. 
Then, sequentially reducing the whole system under consideration $A_0$ by half at each step, after the $k$-th step, the specific entropy assigned to the volume unit $V_0$ takes the form \cite{MICE}:
  \begin{equation}\label{MICE}
    s_0\left(A_0\right)=\frac{H\left(A_0\right)}{V_0}=s_k-\frac 1 2\sum _{j=1}^k\frac{M\left(A_j\right)}{V_j},
  \end{equation}
where $M(A_j)$ is the classical mutual information between identical subsystems $A_j$ with volume $V_j= 2^{-j} V_0$, $s_k=H(A_k)/V_k$ is the specific entropy of the smallest part of the system. In the one-dimensional case we consider, the volume is replaced with the length of a subsystem. For convenience, the value $V_0$ should be a multiple of 2.  

The subsystems under consideration may have different sizes and may not necessarily be next to each other, however for systems with short-range interactions \cite{QI_Sf_Area_laws} one can expect that the mutual information between subsystems $A$ and $B$ scales as their boundary area. In our study splitting the system into halves was done in the way as proposed in the original work \cite{MICE} on the MICE method.  

Previously, it was shown that the MINE, MICE methods can be used for analyzing mutual information and entropy in various physical systems, including classical and quantum systems, under thermal and non-equilibrium conditions \cite{PT_MINE_Kondo, MICE, S_non_eq}. Below we employ both approaches for studying correlation properties of a notable quantum spin model. 

\section*{Results}
In this work we focus on exploring the one-dimensional quantum Ising model with the following Hamiltonian: 
   \begin{equation}\label{Hamiltonian}
    H_{\mathrm{Ising}}=J \sum _{i=1}^N \sigma _i^z \sigma _{i+1}^z - B^x \sum _{i=1}^N \sigma _i^x -B^z \sum _{i=1}^N \sigma _i^z,
   \end{equation}
where $B^z$ and $B^x$ are the longitudinal and transverse magnetic fields, respectively. $\sigma^{x}$ and $\sigma^{z}$ are Pauli matrices. We use the periodic boundary conditions and antiferromagnetic coupling between nearest neighbours, $J=1$. The magnetic field parameters are defined in units of $J$. 

For particular choices of the Ising model parameters the exact solutions are known \cite{Exact}. However, in general case, one needs to solve it numerically, which has already been done by using different methods. For example, the phase structure and magnetic properties of this model have been studied using the density matrix renormalization group (DMRG) \cite{PT_DMRG} and exact diagonalization calculation methods \cite{Degeneracy, Bonfim} as well as variational Monte Carlo using neural networks \cite{PT_NN, PT_MC_long_range, Susceptibility}. The list of physical quantities used for characterizing the system in question includes entanglement entropy, spin-spin correlation functions, quantum fidelity, and others.  

Moreover, the practical importance of using this quantum model as a test object lies in the convenience of its physical implementation in studying the behavior of ultracold atoms and ions in optical traps \cite{PT_cold_atoms, PT_cold_atoms_}, conducting prospective physical experiments \cite{Ry_exp}, numerical modeling of which goes beyond the limits of classical computational resources and testing various ML algorithms \cite{ML_certification}. Thus, the Ising model is of special interest in the fields of condensed matter and quantum computing. Here, we consider this model from the perspective of estimating classical mutual information with a neural network and, on this basis, reconstructing the phase diagram of the Ising model.

\subsection{Estimating classical mutual information with neural network}

To estimate the accuracy of quantifying classical mutual information using neural networks, we consider the solutions of the Ising model, Eq.(\ref{Hamiltonian}) in the range $B^x\in [0.2, 3]$ at the zero longitudinal field $B^z=0$. In this case, according to the previous works \cite{Exact}, there is a quantum phase transition between the antiferromagnetic and paramagnetic phases at $B_{c}^x=1$. To find the ground states of the spin Hamiltonian the SpinED package \cite{SpinED} is used. As it was shown in Ref.\onlinecite{Degeneracy} the accurate description of the degenerate Ising model ground state at weak transverse magnetic fields, $B^x \in [0,0.2)$ requires generating an ensemble of eigenfunctions that belong to degenerate manifold and characterizing their properties, such as quantum correlations, classical mutual information and others. Here we focus on the problem of estimating classical MI in the quantum system and leave the analysis of the degenerate part of the phase diagram to a future investigation.   

\begin{figure}[!h]
    \centering
    \includegraphics[width=0.4771\textwidth]{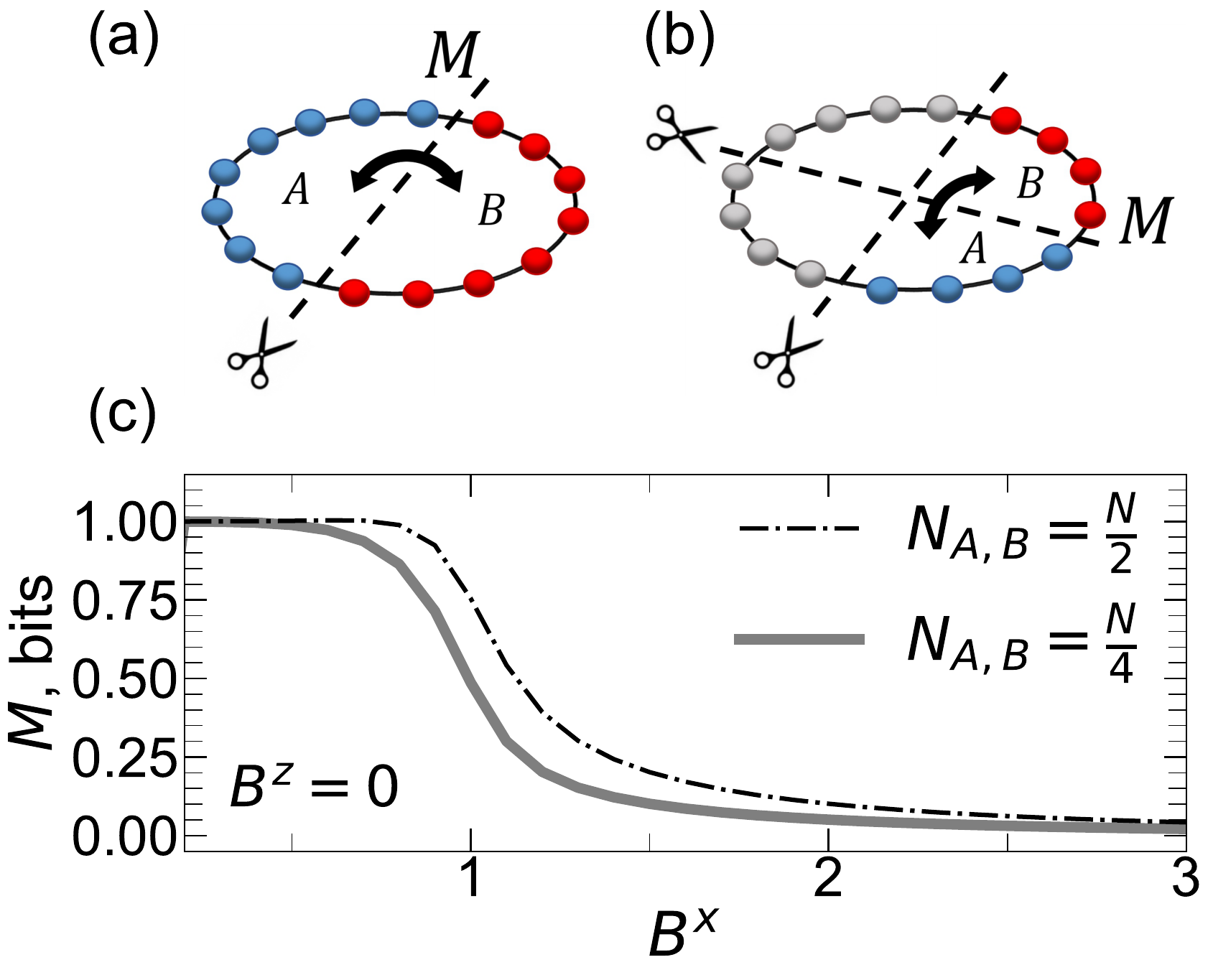}
    \caption{Schematic representation of the system decompositions with $N_A=N_B=\frac{N}{2}$ \textbf{(a)} and $N_A=N_B=\frac{N}{4}$ \textbf{(b)}. Blue and red spheres denote spins from the $A$ and $B$ subsystems, respectively. Gray spheres denote the part of system which is not used in calculations of the mutual information. \textbf{(c)} Exact values of $M(A,B)$ for different decompositions of the system.}\label{FIG2} 
\end{figure}

The calculated ground eigenfunctions of the 16-spin Ising model enable to define the quantities of our interest, which are the Shannon entropy and classical mutual information, Eq.(\ref{MI_def}). Figures \ref{FIG2} \textbf{(a)} and \textbf{(b)} schematically show two variants of the system decomposition we consider. The first one corresponds to the equally-balanced bipartition with $N_{A} = N_{B}=\frac{N}{2}$ (Fig.\ref{FIG2}\textbf{a}). In this case, the subsystems $A$ and $B$ share two common spin pairs. Another type of the decomposition (Fig.\ref{FIG2}\textbf{b}) is characterized by one common spin pair and $N_{A} = N_{B} = \frac{N}{4}$. 

Dependence of the exact values of the classical mutual information on $B^x$ (black curves in Fig.\ref{FIG2}\textbf{c}) reveals a plateau up to the critical field $B_c^{x} = 1$. At $B^{x} > 1$ there is a gradual decrease of $M(A,B)$ with increasing the magnetic field value. Such a behavior can be explained by the structure of the ground state wave function which is mainly characterized by two leading contributions from the basis states $\ket{\uparrow \downarrow \uparrow ... \downarrow}$ and $\ket{\downarrow \uparrow \downarrow ... \uparrow}$ at weak magnetic fields. Increasing the value of $B^x$ leads to delocalization of the ground eigenfunctions in the Hilbert space, the fraction of the basis functions that give non-zero contribution to the ground state increases. In the limit $B^x \rightarrow \infty$ the system is in uniform state and the mutual information is zero.    

\begin{figure}[!h]
    \centering
    \includegraphics[width=0.447\textwidth]{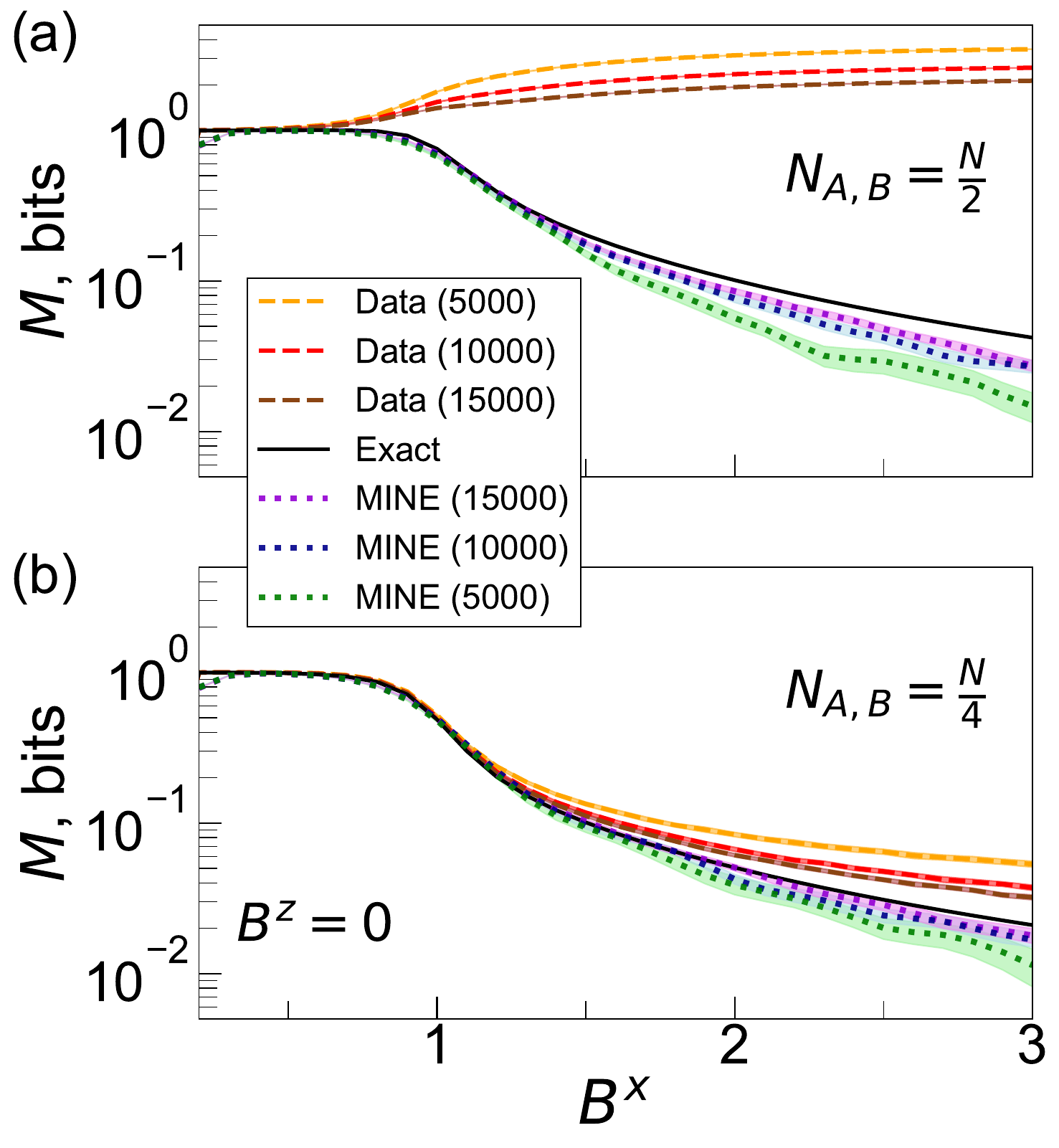}
    \caption{Comparison of the exact values $M$ (black solid line) and approximations obtained with brute-force approach of reconstructing probability distributions, $M_\mathrm{data}$ (orange, red and brown dashed lines) and MINE method, $M_{\boldsymbol{\theta}}$ (green, blue and violet dotted lines). To get the approximate solutions we utilized datasets of 5000, 10000 and 15000 bitstrings. \textbf{(a)} and \textbf{(b)} panels correspond to the results for the system decompositions visualized in Figs.\ref{FIG2} \textbf{(a)} and \textbf{(b)}, respectively. Shaded areas denote standard deviations for the brute-force method values computed on 100 different datasets and the MINE results averaged over 15 neural networks trained independently at each considered magnetic field value.}\label{FIG3} 
\end{figure}

In Fig.\ref{FIG3} we compare the exact values of the classical mutual information, $M(A,B)$, between the subsystems $A$ and $B$ with two approximations $M_{\rm data}$ and $M_{\boldsymbol{\theta}}$ obtained by using a brute-force method for reconstructing state probability and MINE approach, respectively. For these approximations, three sets of 5000, 10000 and 15000 bitstrings were generated from the probability distributions of the ground state wave functions obtained with exact diagonalization. In this way we imitate experimental conditions when one performs a number of projective measurements of a quantum state in the computational $\sigma^z$ basis and use thus obtained bitstrings to characterize underlying wave function. In the case of the brute-force approach, we first use these bit-string sets for reconstructing probability distributions of the subsystem states and then estimate the classical mutual information in the system in question. As for the MINE approach, the bitstrings are directly used for training the neural network that estimates $M_{\boldsymbol{\theta}}(A,B)$ as discussed in the previous section.

\begin{figure*}[t]
    \centering
    \includegraphics[width=0.93\textwidth]{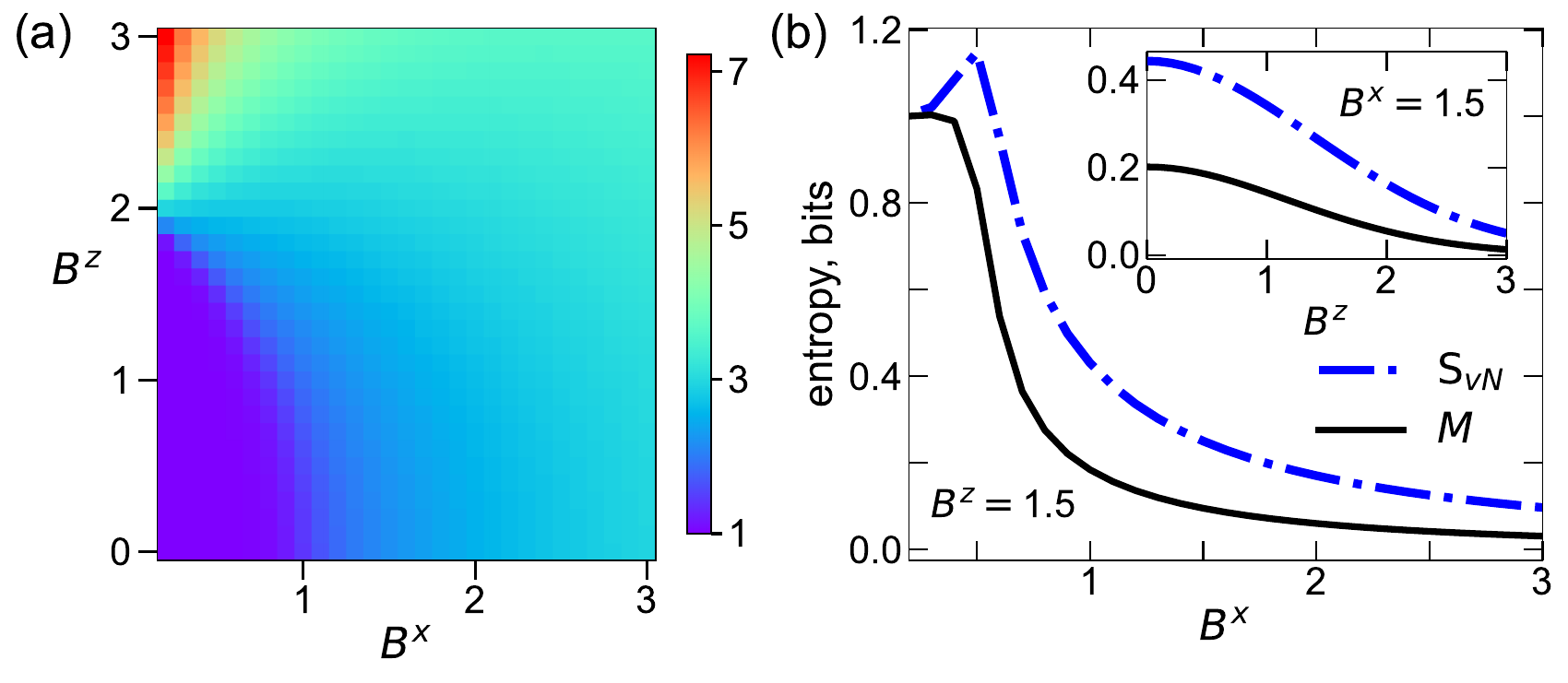}
    \caption{\textbf{(a)} Map that demonstrates the field dependence of the ratio, $\alpha$ between the values of the von Neumann entropy and classical mutual information of the Ising model. These results were obtained for equally-balanced bipartition of the 16-spin system. \textbf{(b)} Examples of magnetic field dependencies for $M$ and $\mathrm{S}_{\rm vN}$ calculated with following field settings: $B^x\in [0.2,3], B^z=1.5$ and $B^x=1.5, B^z\in [0, 3]$ (inset).}\label{FIG4} 
\end{figure*}

First we focus on the MINE results shown in Fig.\ref{FIG3}. The corresponding technical details on the network architecture and training procedure are given in the previous section and Appendix \textbf{A}.  
Within the antiferromagnetic phase ($B^x < 1$) the neural network estimates agree well with the exact classical mutual information for both types of the system decomposition we consider. As the $x$-oriented field increases (measurements are performed in the $\sigma^z$ basis), the system undergoes a paramagnetic state described by a statistical distribution closed to uniform one. Hence, a limited number of bitstrings will be worse at approximating state space at the level of the neural network, which results in more biased estimates of the mutual information \cite{MI_errors}. Put another way, the errors of neural network predictions will increase as $B^{x}$ approaches to 3 (Fig.\ref{FIG3}). Enlarging the training set allows to improve the agreement with the exact values of $M$. Importantly, in both cases of the system decomposition we consider, the MINE results reproduce the exact solution satisfactorily, which is not the case for the brute-force approach discussed below.

Fig.\ref{FIG3}\textbf{a} evidences that in the case of the equally-balanced bipartition ($N_A = N_{B} = \frac{N}{2}$), the difference between the brute-force estimations $M_\mathrm{data}$ and the exact values of mutual information $M$ turns out to be enormous, which is mainly due to the joint entropy contribution in Eq.\ref{MI_def}, $H(A,B)$. To estimate it, we need to calculate the information entropy of the whole system which is described by $2^{16}$ states. Such an amount is not covered by the statistical datasets used in our work, and in this case the advantage of using the MINE method becomes obvious. However, for the second type of the system decomposition ($N_A = N_{B} = \frac{N}{4}$), which requires estimating various probabilities of $2^{8}$ states, the statistics of the generated bitstrings is enough, and the value of the mutual information can be estimated with a small error, which is confirmed with Fig.\ref{FIG3}\textbf{b}. No doubt, when dealing with real-word datasets \cite{MI_benchmark} and real physical systems with a huge number of degrees of freedom and enormous state space, where no exact calculation can be reproduced, this approach will be of little use. Nevertheless, in Appendix \textbf{B} we explore the rate at which the values of $M_\mathrm{data}$ approaches to the exact mutual information $M$.

\subsection{Matching classical mutual information and quantum entanglement entropy}

\begin{figure*}[t]
    \centering
    \includegraphics[width=0.97\textwidth]{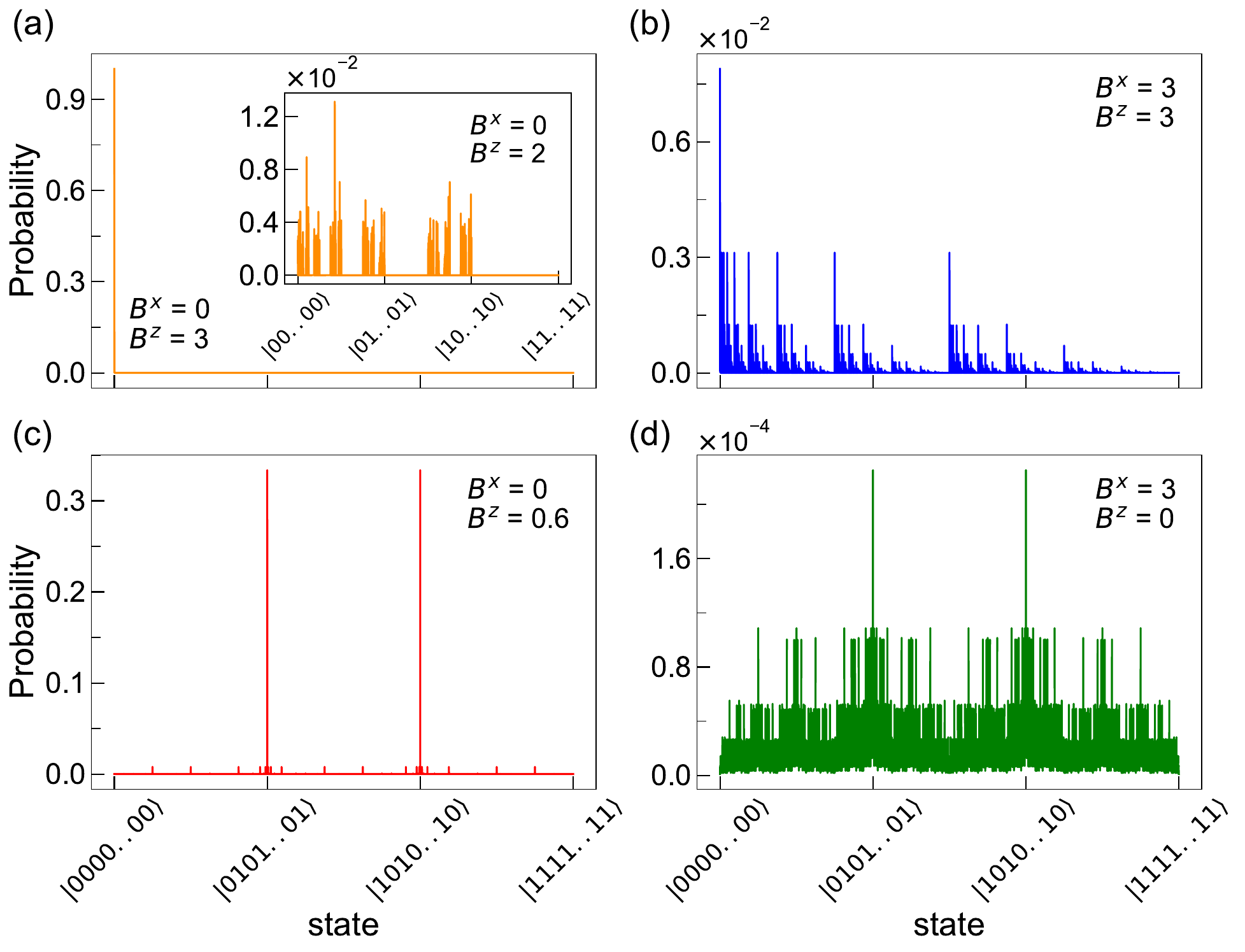}
    \caption{Probability distributions of the different phases: \textbf{(a)} FM-phase ($B^x=0, B^z=3$) (inset figure corresponds to the distribution at the multicritical point $B^x=0, B^z=2$); \textbf{(b)} PM1-phase ($B^x=3, B^z=3$); \textbf{(c)} AFM-phase ($B^x=0.6, B^z=0$); \textbf{(d)} PM2-phase ($B^x=3, B^z=0$)}\label{FIG5} 
\end{figure*}

Possibility to estimate the classical mutual information between different parts of a quantum system with neural networks paves the way to explore other important quantities. Of particular interest in quantum physics is entanglement that can be quantified with the von Neumann entropy, $\mathrm{S}_{\rm vN} (\rho_{A})=-\mathrm{Tr}(\rho_{A} \log_2 \rho_{A})$, where $\rho_{A}$ is the reduced density matrix of the subsystem $A$. In general case, the direct calculation of the entanglement entropy requires reconstruction of the density matrix on a classical computer (quantum tomography) and, therefore, is limited to systems of several qubits, which stimulates developing various measurement-based approaches \cite{Heyl, classical_shadow, toric_code, twin1, twin2, twin3, dissimilarity, Kurmapu, Tiunov, Kharkov} for characterizing quantum systems with different amount of entanglement and estimating $\mathrm{S}_{\rm vN}$. Recently, it was shown that the classical mutual information between halves of a quantum system represents a lower bound for the von Neumann entropy of one of the subsystems \cite{Sf_MI, Sf_MI_}. In other words, one deals with the inequality 
\begin{eqnarray}
M(A,B) \leq \mathrm{S}_{\rm vN} (\rho_{A}),
\end{eqnarray} 
which is intimately related to the Holevo bound \cite{Chuang}.

To elaborate on the correspondence between classical mutual information estimated with the $\sigma^z$-basis measurements and the von Neumann entropy in the case of the Ising model we compute the ratio $\alpha = \frac{\mathrm{S}_{\rm vN} (\rho_{A})}{M(A,B)}$ for $N_{A} = N_{B} = \frac{N}{2}$. These results are presented in Fig.\ref{FIG4}. As one would expect, the antiferromagnetic phase at the weak transverse fields is characterized by $\alpha=1$. Fig.\ref{FIG4}\textbf{b} evidences that increasing longitudinal or transverse fields leads to suppression of both $M(A,B)$ and $\mathrm{S}_{\rm vN} (\rho_{A})$. However, these quantities are characterized by different rates of approaching to zero in the limit of $B^{x}, B^{z} \rightarrow \infty$, which explains large values of the calculated $\alpha$ ratio. Taking into account a weak sensitivity of the considered classical and quantum entropy to the change of the model size (Appendix \textbf{C}) one can use the constructed map (Fig.\ref{FIG4}\textbf{a}) to estimate $\mathrm{S}_{\rm vN} (\rho_{A})$ using the neural network results on $M_{\boldsymbol{\theta}}(A,B)$.

\subsection{Phase diagram}

\begin{figure*}[ht]
    \centering
    \includegraphics[width=1\textwidth]{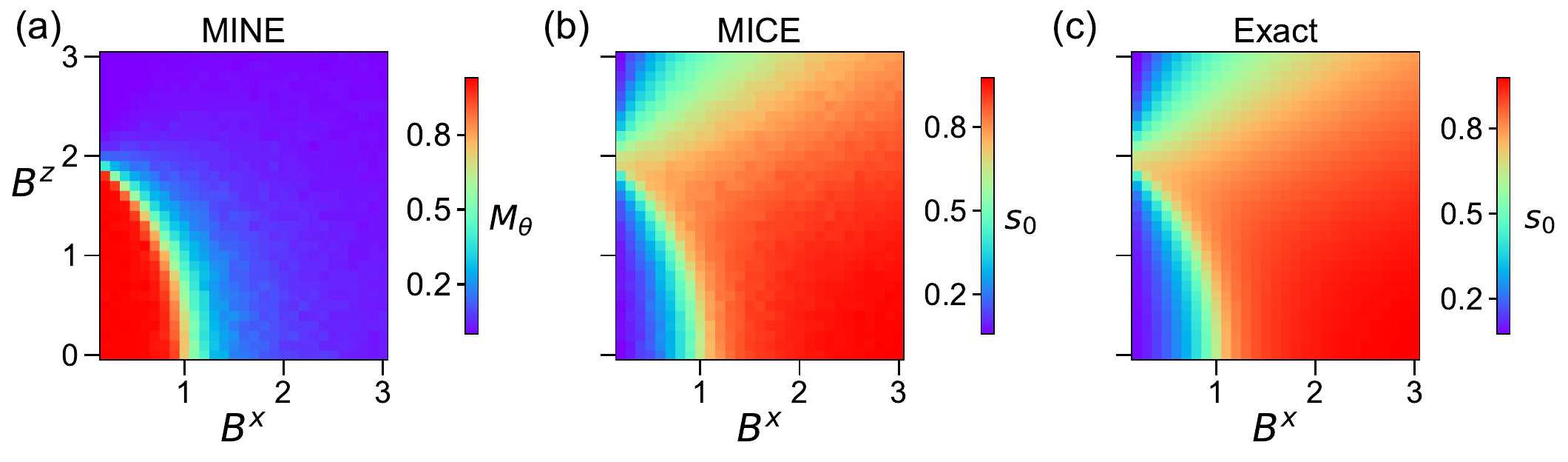}
    \caption{Entropy-based quantities calculated on the $B^x - B^z$ grid with a discrete step of 0.1. $\mathbf{(a)}$  Classical mutual information $M_{\boldsymbol{\theta}}(A,B)$ obtained with MINE approach for equally-balanced bipartition. $\mathbf{(b)}$ The specific entropy quantified with the MICE method.  $\mathbf{(c)}$ Exact values of the specific entropy calculated from the probabilities of the eigenfunction basis states.}\label{FIG6} 
\end{figure*}

Despite simplicity of the Ising model we consider, the reconstruction of its phase diagram in a wide range of the $B^x$ and $B^z$ parameters is still an open problem. To demonstrate the diversity of the quantum states that can be obtained with the Ising Hamiltonian, by the example of Fig.\ref{FIG5} we show the probability distributions of the ground eigenfunctions for some representative points in the parameter space. At $B^z = 3$ and $B^x = 0$ the system is in trivial non-degenerate ferromagnetic state (FM), $\ket{0}^{\otimes 16}$ in the $\sigma^z$ basis and is characterized by the zero classical and quantum entropies. In turn, at weak longitudinal and transverse fields (Fig.\ref{FIG5}\textbf{c}), there is two-fold degenerate antiferromagnetic ground state (AFM) that is described with the entangled wave functions,  $\frac{1}{\sqrt{2}} (\ket{01}^{\otimes 8}+ \ket{10}^{\otimes 8})$ and $\frac{1}{\sqrt{2}} (\ket{01}^{\otimes 8} - \ket{10}^{\otimes 8})$. As we discussed in the previous sections, increasing the transverse field destroys both trivial ferromagnetic and entangled antiferromagnetic states and leads to stabilization of disordered paramagnetic ones. In Figs.\ref{FIG5} \textbf{(b)} and \textbf{(d)} obtained at $B^x$ = 3 one can still recognize the patterns of the pure ($B^x = 0$) AFM and FM states that give the largest contributions to the probability. Importantly, judging by these results the disordered phase is not uniform and, as it was shown in Ref.\cite{Bonfim}, one can distinguish two different paramagnetic regions.   

The classical mutual information quantified with the neural network for the equally-balanced bipartition $N_{A} = N_{B} = \frac{N}{2}$ and presented in Fig.\ref{FIG6}\textbf{a} clearly shows AFM (red area) and paramagnetic (violet area) phases on the $B^x - B^z$ plane. Since both ferromagnetic and paramagnetic states feature zero correlation between different parts of the whole system, they cannot be distinguished with the $M_{\boldsymbol{\theta}}(A,B)$ quantity.
However, as can be seen from Fig.\ref{FIG6}\textbf{b} the specific entropy reconstructed using Eq.\ref{MICE} reveals two transition areas, AFM-PM ($B^z \in [0,2)$) and FM-PM ($B^z \in (2,3]$). In the case of the pure AFM state characterized by two equiprobable contributions, $\ket{01}^{\otimes8}$ and $\ket{10}^{\otimes 8}$, the total Shannon entropy is equal to 1 and the corresponding specific entropy value approaches to zero as the number of spins increases, $s_{0} = \frac{1}{N}$. The FM value of $s_0$ is zero. A high quality of reconstructing $s_0$ with neural networks is confirmed by the comparison with exact calculations presented in Fig.\ref{FIG6}\textbf{c}.  

\begin{figure}[h!]
    \centering
    \includegraphics[width=0.465\textwidth]{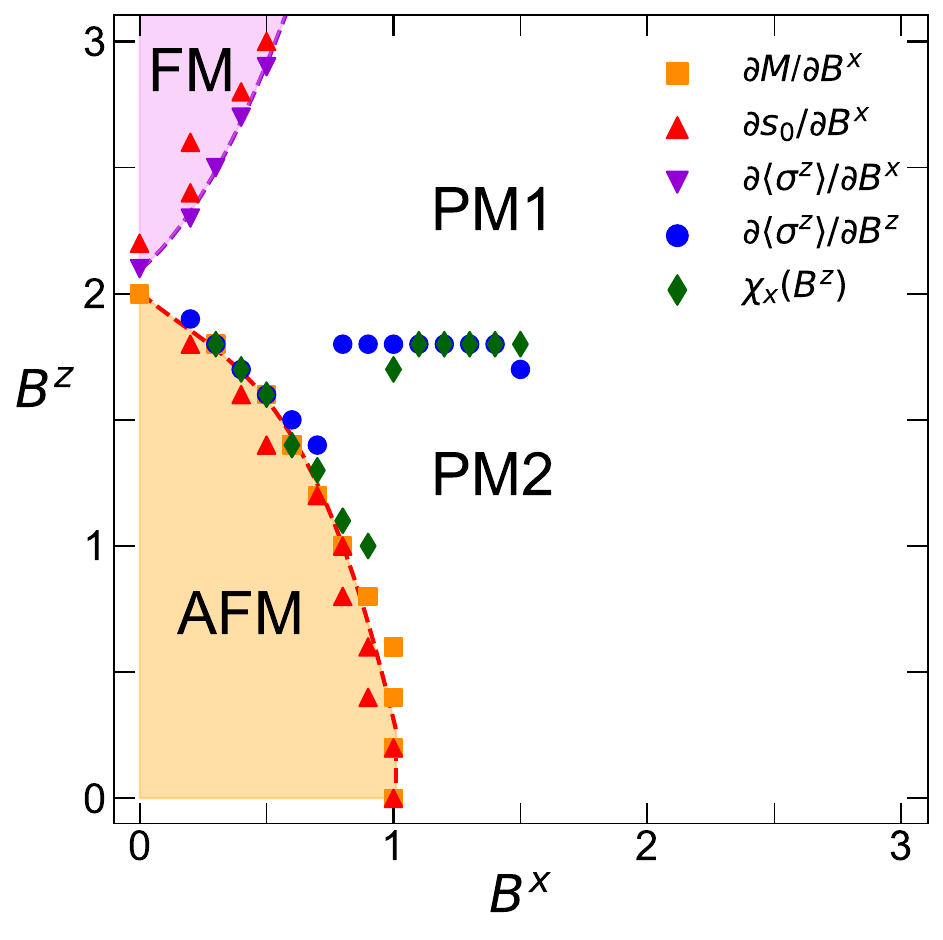}
    \caption{Phase diagram of the Ising model in the longitudinal and transverse fields reconstructed with different quantities. They include magnetic field derivatives of the classical mutual information (orange squares), specific entropy (red triangles), average values of the z-oriented spin operator (violet triangles and blue circles) as well as fidelity susceptibility (green rhombuses).}\label{FIG7} 
\end{figure}

In general case reconstructing the phase diagram assumes detecting boundaries between different states of the system in question. For these purposes we first take derivatives of the classical mutual information and specific entropy discussed above with respect to the transverse magnetic fields $B^x$. From Fig.\ref{FIG7} one can see that the extrema of these entropy-based quantities allow to reveal the boundaries between trivial FM and PM states as well as entangled AFM and paramagnetic phases. The AFM-PM transition is in good agreement with previously reported magnetic phase diagrams of the Ising model \cite{PT_DMRG,Bonfim} constructed with other measures. At the same time, in literature we didn't find an information concerning FM-PM transition which takes place at high longitudinal magnetic fields. The observed fluctuations of $\frac{\partial M}{\partial B^x}$ and $\frac{\partial s_0}{\partial B^x}$ around the ideal phase boundaries can be explained by the size of the magnetic field grid characterized by the step of 0.1 between nearest points. Another source for discrepancy is related to the fact that we use the neural network approach which provides approximate values of the classical mutual information.  

\begin{figure}[h!]
    \centering
    \includegraphics[width=0.48\textwidth]{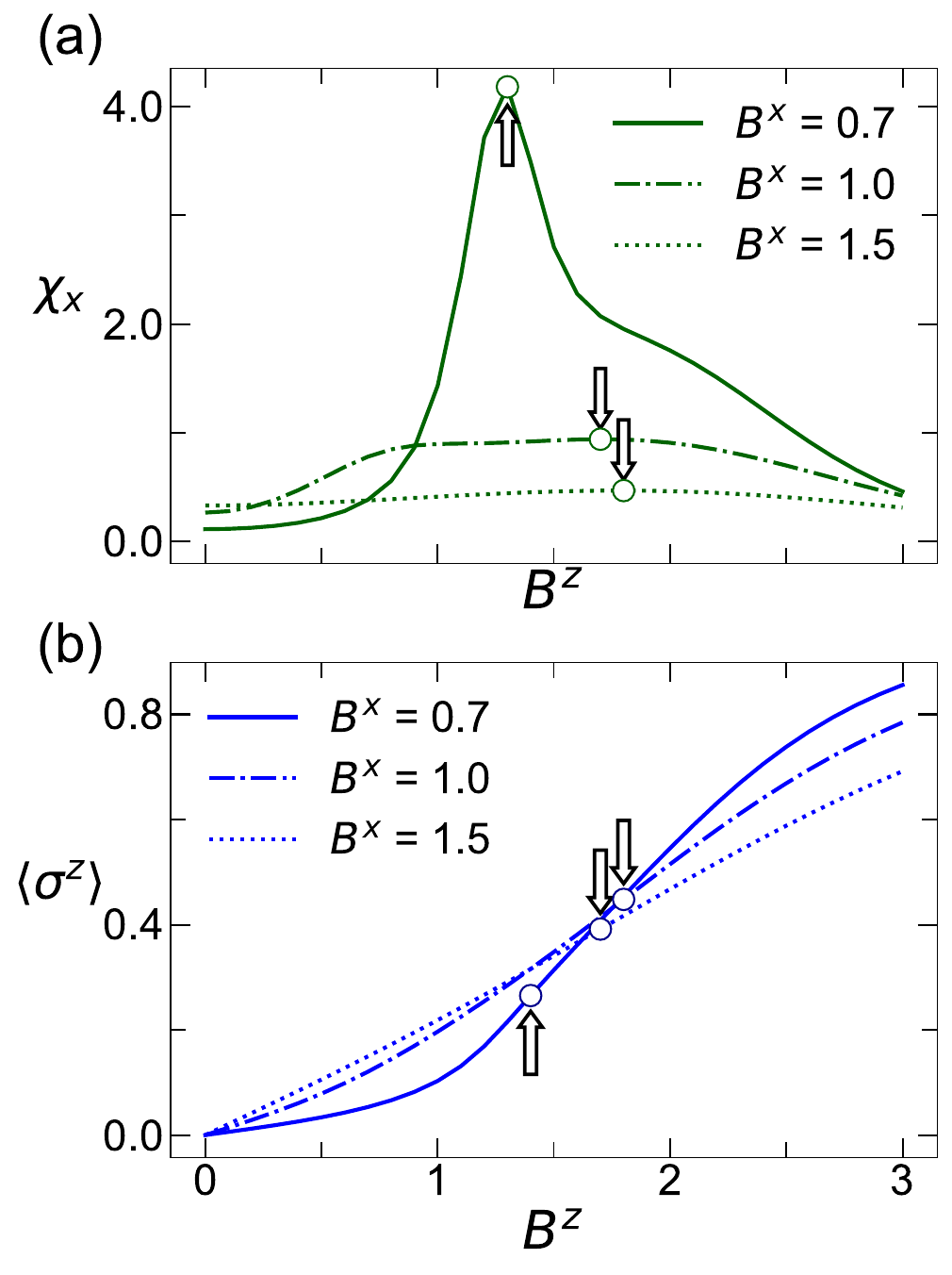}
    \caption{\textbf{(a)} Fidelity susceptibility calculated with Eq.\ref{Chi} for different values of the transverse magnetic field. Arrows indicate the susceptibility maxima. \textbf{(b)} Field dependence of the average values of the local z-oriented spin operator for the same $B^{x}$ as in \textbf{(a)}. Arrows indicate the maxima of the magnetization derivatives.}\label{FIG8} 
\end{figure}

To complete our consideration of the Ising model phase diagram we have employed  a fidelity susceptibility that allows unsupervised detection of the quantum transitions \cite{Susceptibility_}. Such a quantity is given by
 \begin{equation}\label{Chi}
     \chi_{\nu }\left(B^{\mu}\right)=\frac{2\left(1-F\left(B^{\mu} ,{\delta}\right)\right)}{{\delta}^2}+O({\delta}^3)
   \end{equation}
with the fidelity for the fixed value of a magnetic field $B^{\nu}$
 \begin{equation}\label{Fidelity_square}
     F\left(B^{\mu} ,{\delta}\right)=\left|\left.\left\langle \psi (B^{\mu} )\right.\right|\left.\psi (B^{\mu}
     +{\delta})\right\rangle \right|,
   \end{equation}
where $\delta$ is a small shift of the magnetic field value, $\mu = z$ and $\nu=x$ or vice versa. Following Ref.\onlinecite{Bonfim} we used $\delta=0.001$. In this approach quantum phase transitions are detected through the extrema of the fidelity susceptibility.
In the case of the Ising model with longitudinal and transverse magnetic fields the authors of Ref.[\onlinecite{Bonfim}] have calculated $\chi_{\nu }\left(B^{\mu}\right)$ in the range $B^x \in [0,2]$ and $B^z \in [0,2]$. They have revealed not only the  AFM-PM transition but also a critical area inside the paramagnetic phase that is related to balance between ferromagnetic $\ket{0}^{\otimes 16}$ and antiferromagnetic $\ket{01}^{\otimes8} + \ket{10}^{\otimes 8}$ contributions to the quantum state of the system in the paramagnetic phase.  In Figs.\ref{FIG7} and \ref{FIG8}\textbf{a} we reproduce the result of Ref.\onlinecite{Bonfim} for $\chi_{x}\left(B^{z}\right)$ (green rhombuses) by the example of the 16-spin system. It is important to note that for the given $B^x$ the $\chi_{x }\left(B^{z}\right)$ function becomes more and more flat as the longitudinal magnetic field $B^z$ increases and the magnitude of its maximum decreases (Fig.\ref{FIG8}\textbf{a}). Based on this we conclude that distinguishing different paramagnetic regions with the fidelity susceptibility becomes impossible for $B^x > 1.5$.  

Since in the case of large scale quantum systems the calculation of $\chi_{\nu }\left(B^{\mu}\right)$ could be a challenging task \cite{Susceptibility, Susceptibility_, Fidelity}, we have examined a local spin correlation function, $\braket{\sigma^z}$ that is traditionally used as a order parameter for Ising model and is easily accessible in experiments. Such a correlation function is zero in the antiferromagnetic phase and reaches the maximal value of 1 for the ferromagnetic state. Thus, this quantity can also capture the balance between contributions of ferro- and antiferromagnetic basis states. As follows from Fig.\ref{FIG8}\textbf{b} $\braket{\sigma^z}$ as a function of $B^z$ at $B^x =0.7$ is characterized by a change in slope at about 1.4 and generally becomes more linear as $B^x$ increases. In the range $B^x \in [1,1.5]$ the maximum of the magnetization derivative with respect to $B^z$ behaves similar to fidelity susceptibility. In turn, the calculations of $\frac{\partial \braket{\sigma^z}}{\partial B^x}$ at fixed longitudinal magnetic fields reveal the boundary between FM and PM states, which agrees with results of the specific entropy simulations. 

\section*{Conclusions}
To sum up, we have employed a neural network approach for estimating classical mutual information to characterize correlations in a quantum system. For that a limited number of bitstrings obtained from the projective measurements in the $\sigma^z$ basis was used. The accuracy of the neural network estimates was demonstrated by the example of the Ising model that hosts entangled antiferromagnetic and paramagnetic states. In addition, we have explored the connection between quantum entanglement and classical mutual information, which can be important for theoretical description of the large scale experimental simulation of the Ising Hamiltonian. Using classical mutual information and other quantities, the phase diagram of the considered quantum system was reconstructed with special attention paid to the analysis of paramagnetic states. This Ising model example shows that a reliable estimation of the entropy-based quantities with neural networks is possible even when there is deficit of information.    

In addition to the task of constructing phase diagrams one can consider another direction of practical importance that is certification of quantum states and quantum devices. For instance, estimating classical mutual information with limited number of bitstrings can be used for certifying complex highly-entangled states such as Haar-random states \cite{Haar1,Haar2}. In general, characterization of such states delocalized in the Hilbert space requires performing measurements in different bases \cite{dissimilarity} and their optimal choice is still an open problem. Benchmarking regimes of interferometers (boson sampling problem) \cite{boson2, boson3} likewise attracts considerable attention and stimulates developing different numerical approaches \cite{Boson_correlation, Boson_Rubtsov1, Boson_Rubtsov2, Boson_Rubtsov3}. In principle, the classical mutual information calculated exactly enables discriminating distinguishable and indistinguishable regimes of boson samplers, as shown in Ref.\onlinecite{Boson_Hamming}. However, at this moment it is not clear whether such a benchmarking is possible with approximate values of $M_{\boldsymbol{\theta}}(A,B)$.  

\section*{Acknowledgments}
This work was supported by the Ministry of Science and Higher Education of the Russian Federation (theme FEUZ-2023-0013).

\section{Appendix}

\subsection{Technical details}
In this study, the MINE calculations were performed by using the package \cite{MINE_code}. First, this realization of the MINE approach has been tested by us with reproducing the results of Ref.\cite{MICE}. Then, in the case of the quantum Ising model we focus on, a fully connected neural network with input, 3 hidden and output layers was used. Input layer contains 16 neurons, which corresponds to the number of spins in the Ising model. Each hidden layer contains 64 neurons with the ReLU activation function. The output layer consists of one neuron that defines the value of the $f_{\boldsymbol{\theta}}$ function in Eq.\ref{MI_D_KL} for the given spin state (bitstring). The weights were updated according to the gradient backpropagation algorithm with the SGD optimizer. The learning rate and momentum were equal to 0.01 and 0.8, respectively. The dropout value of 0.1 was used. 

\begin{figure}[!h]
    \centering\includegraphics[width=0.48\textwidth]{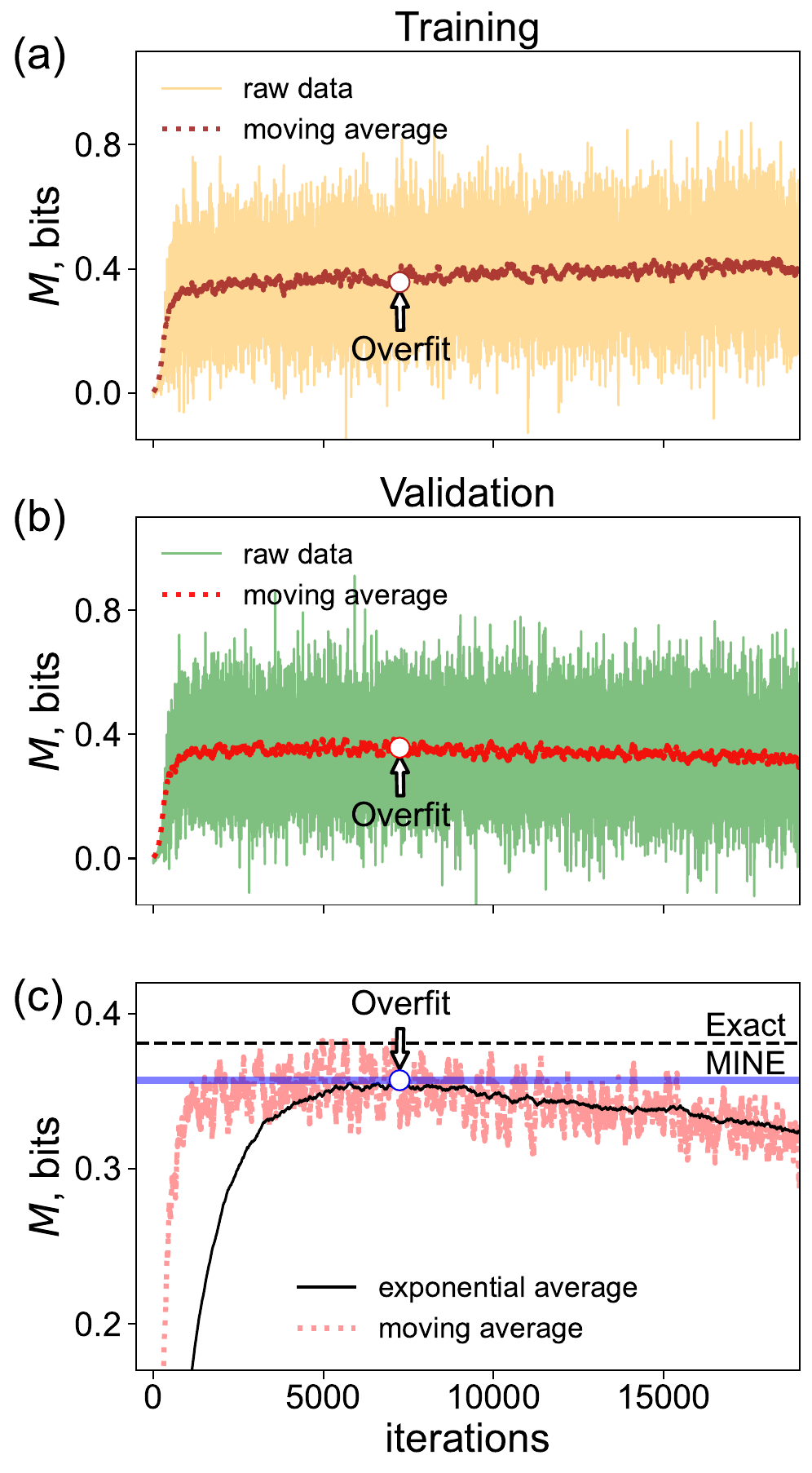}
    \caption{Example that demonstrates training a neural network with the data obtained for $B^x=1$ and $B^z=1$. \textbf{(a)} Values of the classical mutual information estimated with training set at different iterations. Yellow and brown lines denote raw and average values of MI. \textbf{(b)} MI values estimated with the validation set.  All curves in (a) and (b) are processed by a smoothing filter (MA) with equally probable neighbor iterations. \textbf{(c)} Comparison of results obtained with moving average (red dotted line) and exponential average, Eq.\ref{EMA} (black solid line) filters for raw data presented in (\textbf{b}). The resulting mutual information (blue line) is determined by us as the last maximum value obtained at a certain validation step, after which the neural network begins to overfit.}\label{FIG9} 
\end{figure}

Fig.\ref{FIG9} gives example of main stages of the learning procedure we use. At first, a complete set of spin states that includes 15000 bitstrings is split in a ratio of 80$\%$ to 20$\%$. Then, the most bitstrings are employed to train the neural network (Fig.\ref{FIG9}\textbf{a}), and another smaller part (the validation set) is used to control overfitting (Fig.\ref{FIG9}\textbf{b}), and find an approximate value of the classical mutual information (Fig.\ref{FIG9}\textbf{c}). At each iteration, evaluation of contributions to $M_{\boldsymbol{\theta}}(A,B)$ is performed by using a batch of 256 bitstrings randomly taken from the training set. 
Based on the qualitative change in the behavior during the validation stage, it is possible to find the iteration at which  the neural network starts to overfit. In the case of the MINE algorithm, the inflection point is the maximum of the test curve, so it is taken as the estimated value of $M_{\boldsymbol{\theta}} (A,B)$ (Fig.\ref{FIG9}\textbf{c}). The total number of iterations we use is 20000, but the training process can be stopped earlier depending on the validation results, as discussed above.

To get an accurate prediction of the classical mutual information, it is necessary to smooth the output values of the neural network (raw data) that are characterized by strong fluctuations due to the small size of the bit-string batch. For that we used the usual moving average (MA) and exponential average (EMA), as suggested in Ref.\cite{MICE}:
  \begin{equation}\label{EMA}
    \langle M\rangle_{\mathrm{EMA}}^{(i+1)}= \langle M\rangle_{\mathrm{EMA}}^{(i)}+\gamma(M_{\boldsymbol{\theta}}^{(i)}-\langle M\rangle_{\mathrm{EMA}}^{(i)}).  
  \end{equation}
Here $\gamma$ controls smoothing (it is set to 0.001), $M_{\boldsymbol{\theta}}^{(i)}$ is the mutual information value estimated with neural network at the $i$-th iteration.

\subsection{Approximating $M_{\mathrm{data}}$ with a limited number of bitstrings} 

\begin{figure}[!h]
    \centering
    \includegraphics[width=0.48\textwidth]{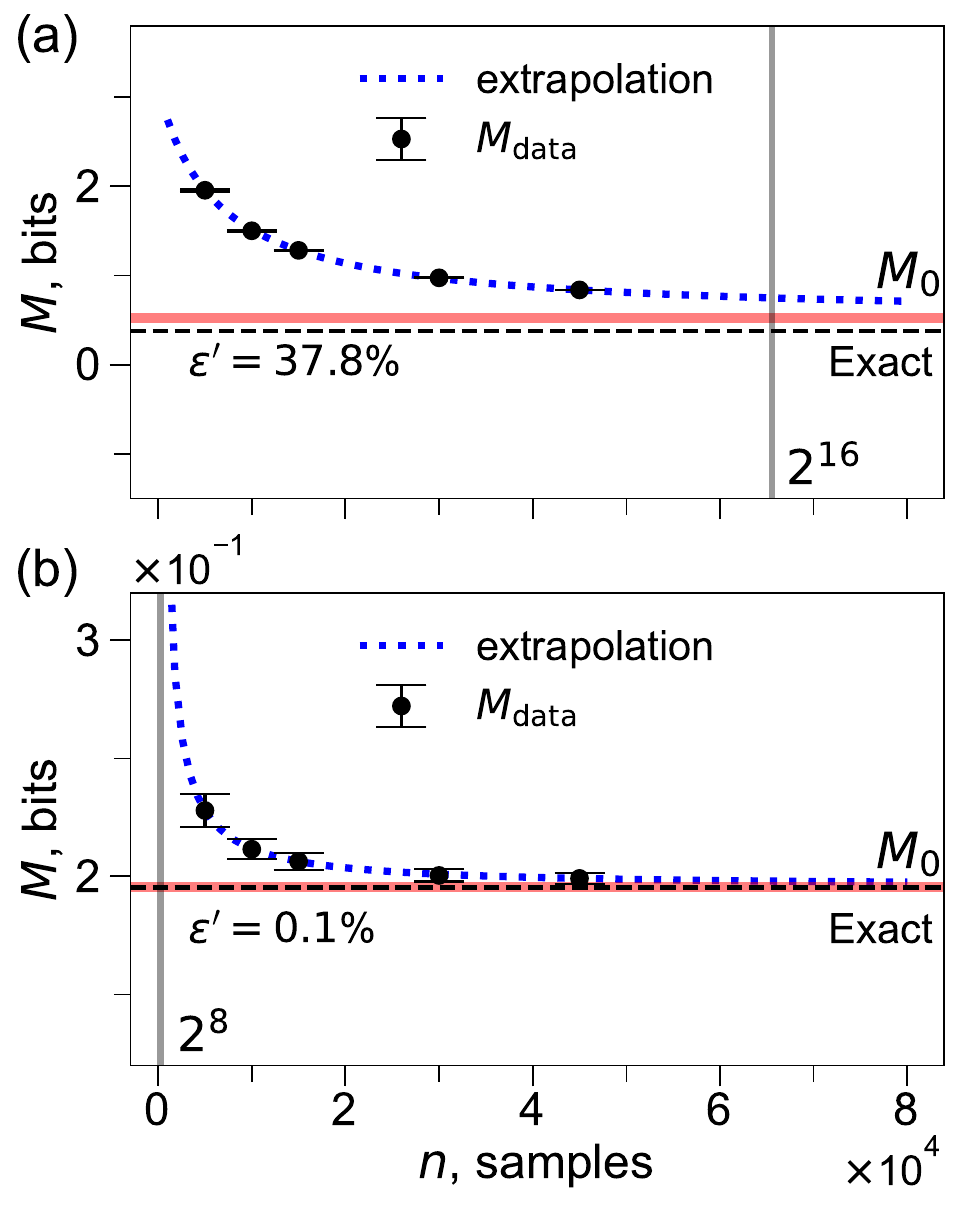}
    \caption{Convergence of $M_{\mathrm{data}}$ at increasing the number of samples $n$ used for estimating the probabilities of the states of the Ising model with $B^x=1$ and $B^z=1$. Blue dotted lines denote the resulting MI functions that fit MI estimates (black circles).  $\textbf{(a)}$ and $\textbf{(b)}$ correspond to the $N_{\rm A,B} = \frac{N}{2}$ and $N_{\rm A,B} = \frac{N}{4}$ partitions, respectively. Gray vertical lines denote the largest state space dimensions, $2^{16}$ (a) and $2^{8}$ (b)  of the subsystems used for estimating the classical mutual information. Error bars denote standard deviations obtained at averaging over 100 different datasets. $\varepsilon^{\prime}=|M_0-M|/M$ stands for the relative error between asymptotic upper bound (red line) and exact value of the mutual information (black dashed line).}\label{FIG10} 
\end{figure}

In this section we explore quality of reconstructing the classical mutual information with brute-force approach aimed at approximating probabilities of different system's states with finite number of samples. For that we start with the example of the 16-spin Ising model at $B^x=1$ and $B^z=1$, whose ground state  belongs to paramagnetic phase (Fig.\ref{FIG7}). Fig.\ref{FIG10} gives the $M_{\rm data}$ estimates obtained with 5000, 10000, 15000, 30000, and 45000 samples for the $N_{\rm A,B} = \frac{N}{2}$ and $N_{\rm A,B} = \frac{N}{4}$ partitions presented in Figs.\ref{FIG2} \textbf{a} and \textbf{b}. 

Using the least squares method the obtained data were fitted with the following MI function
\begin{equation}\label{tendency}
       {M_\mathrm{data}(n)=M_0 +\frac{k}{n-n_0}},
\end{equation}    
where $n$ is the number of samples, $k$ is a constant, $M_0$ and $n_0$ are asymptotic bounds. The former asymptote corresponds to the estimated value of the mutual information in the limit of the infinite number of samples, and the latter defines a shift along the sampling axis to provide the best approximation with Eq.\ref{tendency}.

The relative error $\varepsilon'$ in estimating $M$ for the $N_A=N_B=\frac{N}{2}$ partition is equal to 37.8 $\%$, which is two orders of magnitude greater than $\varepsilon' = 10^{-1}$ $\%$ for the case of the second partition we consider with $N_A=N_B=\frac{N}{4}$. Such a difference can be explained by the fact that in the first case, all points used for fitting the MI function were obtained with the number of samples smaller than the dimension of the state space, $2^{16}$. For the second partition, all the considered values of $n$ are larger than the size of the probability distribution of $2^{8}$ that describes largest subsystem. This results in a better agreement of the $M_0$ asymptote with the exact value of mutual information.

The brute-force estimation of mutual information with Eq.\ref{tendency} has an inversely proportional dependence on the number of samples $M_{\mathrm{data}}(n) \sim \frac{1}{n}$, and in the limit $n\to \infty$ this expression is simply replaced by $M_{\mathrm{data}}=M_{0}$. Knowing the function along which the convergence to the exact bound of $M$ is carried out, predictions about the mutual information at several points on this curve can be made by selecting the most appropriate coefficients $M_0$ and $k$ with some certain limits of their error, which are mainly determined by the size of the state space of the considered system.

With the growth of the number of spins in the system in question, this approach becomes irrational to use because of the need to perform a significant amount of calculations. However, it is still interesting to explore, in general case, the rate at which the values of $M_\mathrm{data}$ tend to the upper bound of meaning $M$, because despite the many methods developed to better approximate the lower bound of this characteristic, it is likely that there is no such method that would not be subject to bias and could process even extremely noisy data \cite{MI_errors}.

\subsection{Size scaling of the ratio between $M$ and $\mathrm{S}_{vN}$}

\begin{figure*}[]
    \centering
    \includegraphics[width=0.92\textwidth]{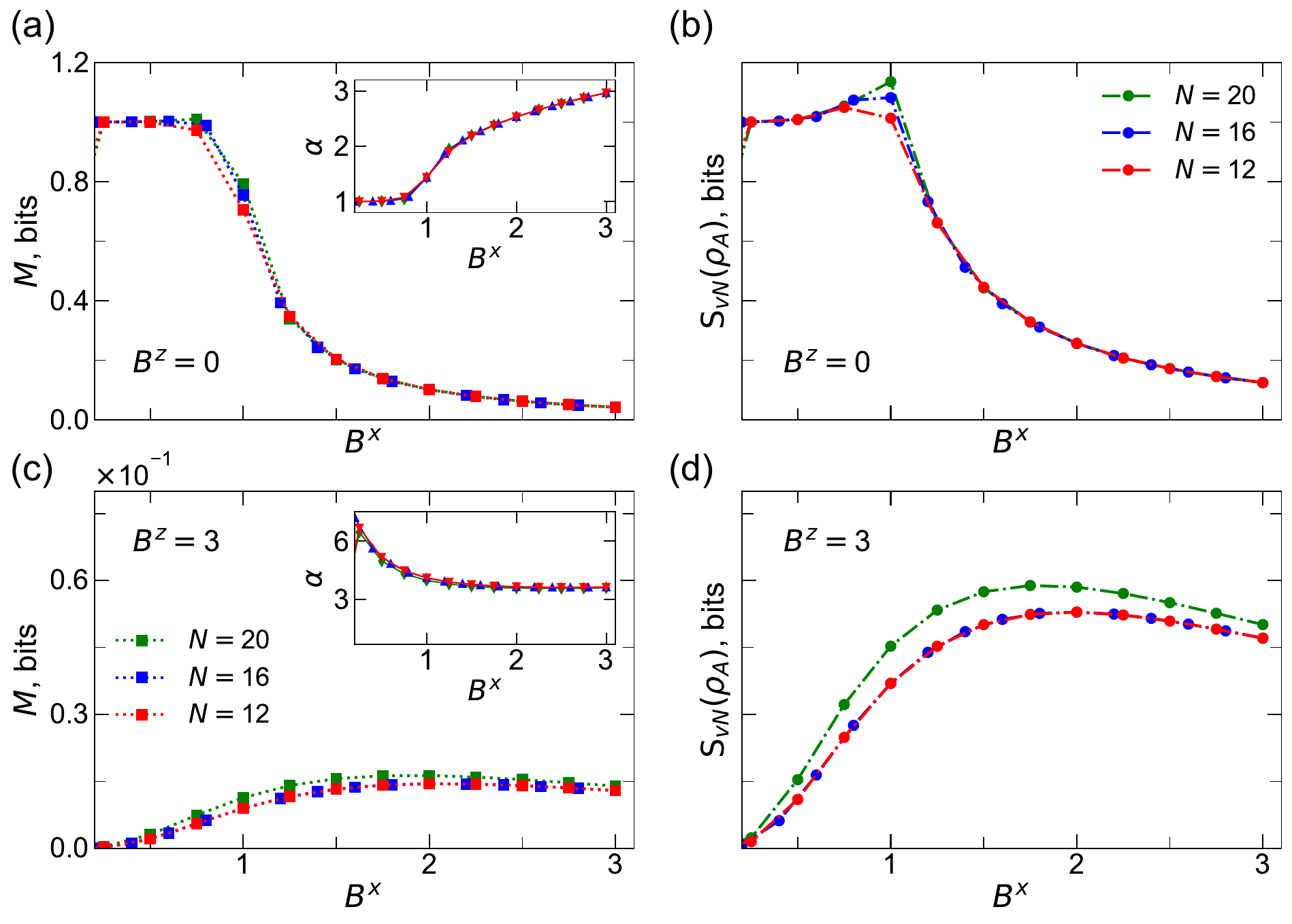}
    \caption{Size scaling of the exact classical mutual information \textbf{(a), (c)} and entanglement entropy \textbf{(b), (d)}. Results were obtained for the equally-balanced bipartition of spin systems of different sizes $N$=12, 16 and 20. The ratio $\alpha$ between the classical mutual information $M$ and the von Neumann entropy $\mathrm{S}_{\rm vN}$ is shown in the inset figures for both cases of parameter range, i.e. $B^x\in [0.2,3], B^z=0$ \textbf{(a), (b)} and $B^z=3$ \textbf{(c), (d)}.}\label{FIG12} 
\end{figure*}

In this section we explore sensitivity of the classical mutual information and von Neumann entropy to increasing the system's size, $N$. Figure \ref{FIG12} shows nearly identical behavior of both quantities calculated exactly by using the ground eigenstates obtained from exact diagonalization at different values of $N$.
Having fixed the longitudinal field at $B^z = 0$ we observe a peak of the von Neumann entropy at the critical transverse field $B^x = 1$ whose value increases when $N$ becomes larger. This is consistent with the non-analytical behavior of the entanglement entropy in the region of quantum phase transitions \cite{Entanglement_Many_Body_systems, PT_Sf}. At the same time the size dependence of the classical mutual information is less pronounced compared with $S_{\rm vN}$. Increasing the longitudinal field leads to a considerable suppression of the $M(A,B)$ and $S_{\rm vN}$ values, for instance, when one reaches $B^z =3$ they become one order of magnitude smaller then that for $B^z=0$. Importantly, the ratio $\alpha = \frac{S_{\rm vN}(\rho_{\rm A})}{M(A,B)}$ is robust with respect to the change of the system's size (Fig.\ref{FIG12} insets). It means that the numerical map shown in Fig.\ref{FIG4}\textbf{a} in the main text can be used to estimate the amount of quantum entanglement in a physical system of any size with the classical mutual information approximated by using the MINE method with a limited number of experimental measurements performed in the computational basis.

\end{document}